\documentclass[twocolumn,pre,superscriptaddress]{revtex4}   

\usepackage{dcolumn}
\usepackage{amsmath}

\usepackage{graphicx}

\newcommand{\etal}{{\it{}et~al.}}
\newcommand{\defn}{\textit}

\begin{document}

\title{Aspirational pursuit of mates in online dating markets}

\author{Elizabeth E. Bruch}
\affiliation{Department of Sociology, University of Michigan, Ann Arbor, Michigan, USA}
\affiliation{Center for the Study of Complex Systems, University of Michigan, Ann Arbor, Michigan, USA}

\author{M. E. J. Newman}
\affiliation{Center for the Study of Complex Systems, University of Michigan, Ann Arbor, Michigan, USA}
\affiliation{Department of Physics, University of Michigan, Ann Arbor, Michigan, USA}

\begin{abstract}
  Romantic courtship is often described as taking place in a dating market where men and women compete for mates, but the detailed structure and dynamics of dating markets have historically been difficult to quantify for lack of suitable data.  In recent years, however, the advent and vigorous growth of the online dating industry has provided a rich new source of information on mate pursuit. Here we present an empirical analysis of heterosexual dating markets in four large US cities using data from a popular, free online dating service.  We show that competition for mates creates a pronounced hierarchy of desirability that correlates strongly with user demographics and is remarkably consistent across cities.  We find that both men and women pursue partners who are on average about 25\% more desirable than themselves by our measures and that they use different messaging strategies with partners of different desirability.  We also find that the probability of receiving a response to an advance drops markedly with increasing difference in desirability between the pursuer and the pursued.  Strategic behaviors can improve one's chances of attracting a more desirable mate, though the effects are modest.
\end{abstract}

\maketitle

\section{Introduction}
It is a common observation that marriage or dating partners strongly resemble one another in terms of age, education, physical attractiveness, attitudes, and a host of other characteristics~\cite{schwartz2013}. One possible explanation for this is the \defn{matching} hypothesis, which suggests that men and women pursue partners who resemble themselves. This in turn implies that people differ in their opinions about what constitutes a desirable partner or at least about who is worth pursuing.  At the other extreme, and more in line with biological studies of mate selection~\cite{andersson1994,darwin1874,wong2005}, lies the \defn{competition} hypothesis, which assumes that there is consensus about what constitutes a desirable partner and that mate seekers, regardless of their own qualifications, pursue those partners who are universally recognized as most desirable~\cite{becker1973,mortensen1988,gale1962,laumann2004}.  Paradoxically this can also produce couples who resemble one another in terms of desirability, as the most desirable partners pair off with one another, followed by the next most desirable, and so on.  To the extent that desirability correlates with attributes like age, physical attractiveness, and education, the matching and competition hypotheses can, as a result, produce similar equilibrium patterns of mixing~\cite{becker1973,Kalick1986,Xie2015}.

However, while the two hypotheses may produce similar outcomes, they carry very different implications about the processes by which people identify and attract partners.  If there is consensus about who is desirable, it creates a hierarchy of desirability~\cite{walster1966,taylor2011,HHA10} such that individuals can, at least in principle, be ranked from least to most desirable and their ranking will predict how and to what extent they are pursued by others.  Historically, however, such hierarchies have been difficult to quantify.  Since they reflect which partners people pursue, and not just who people end up with, one would need a way to observe unrequited overtures as well as requited ones in order to determine who people find desirable.  Online dating provides us with an unprecedented opportunity to observe both requited and unrequited overtures at the scale of entire populations.

As data from online dating web sites have become available, a number of studies have explored the ways in which mate choice observed online can inform the debate about matching versus competition.  These studies typically focus on how specific attributes of individuals shape their browsing and messaging behavior.  The results indicate that with respect to attributes such as physical attractiveness and income, people tend to pursue the most attractive partners~\cite{HHA10,lee2008,walster1966}, while for other attributes, such as race/ethnicity or education, the overwhelming tendency is to seek out someone similar~\cite{lewis2013,anderson2014}. Thus, people compete on some attributes and match on others.  While these studies provide valuable insights about matching and competition on an attribute-by-attribute basis, however, they do not capture the overall dating hierarchy that reflects total demand for each person in the market.

In this paper we report results from a quantitative study of aspirational mate pursuit in adult heterosexual romantic relationship markets in the United States, using large-scale messaging data from a popular online dating site (\textit{Materials and Methods}, Data).  We provide a crisp, operational definition of desirability that allows us to quantify the dating hierarchy and measure, for instance, how far up that hierarchy men and women reach for partners and how reach is associated with the likelihood of getting a response.  We also explore the ways in which people tailor their messaging strategies and message content based on the desirability of potential partners, and how desirability and dating strategy vary across demographic groups.

\section{Results}

\begin{figure}
\begin{center}
\includegraphics[width=\columnwidth]{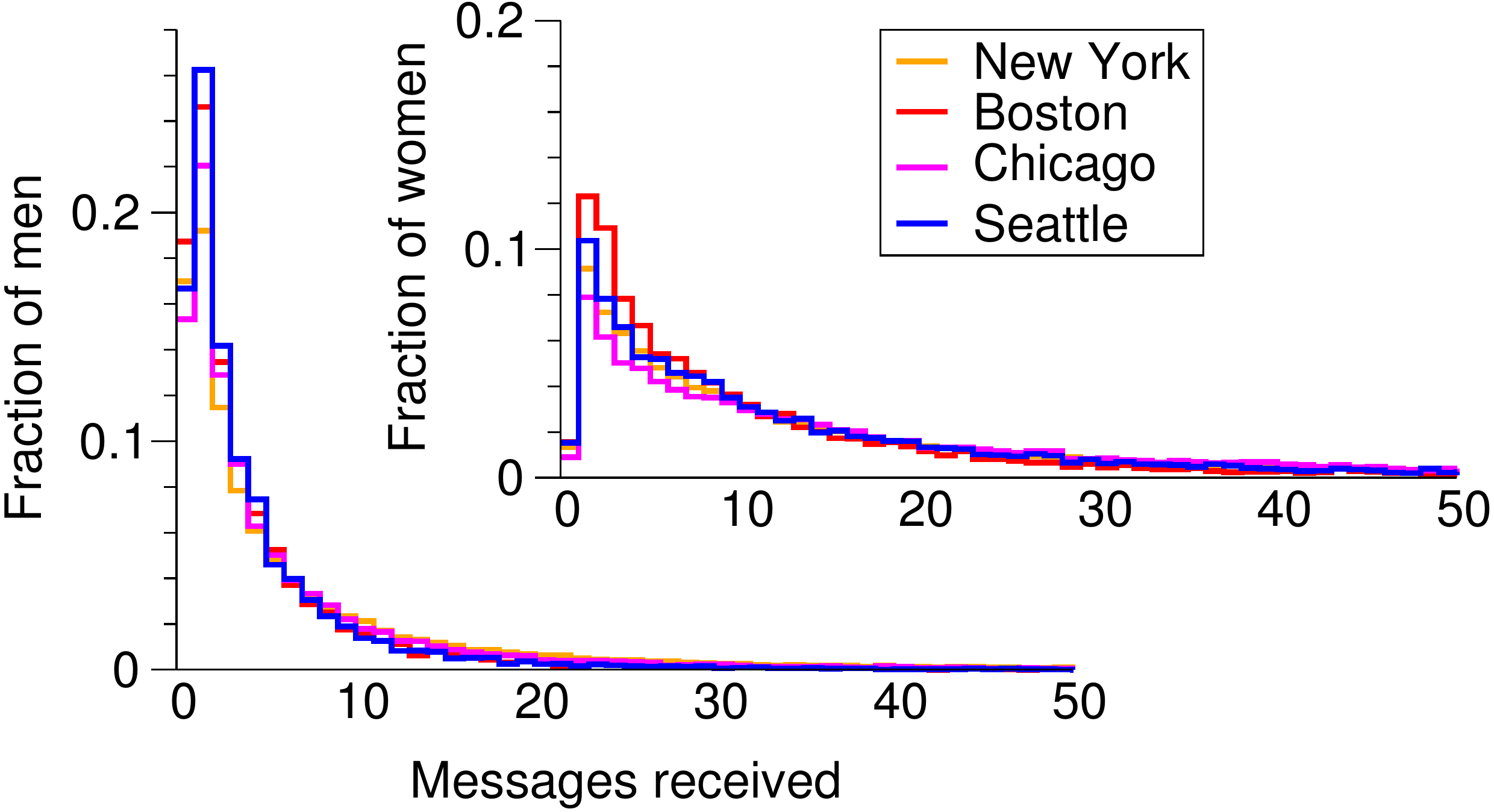}
\end{center}
\caption{Histograms of the number of first messages received by men and women in each of our four cities.}
\label{fig:hierarchy}
\end{figure}

To study individual desirability we focus on messages between users of the web site over a 1-month period in four cities: New York, Boston, Chicago, and Seattle.  At the simplest level, one can quantify desirability by the number of messages a user receives, and specifically the number of \emph{initial} messages, since it is the first contact between a pair of individuals that most reliably indicates who finds whom attractive.  Figure~\ref{fig:hierarchy} shows the distribution of this quantity separately for men and women in each of the cities.  The distribution is roughly consistent across cities and, though women receive more messages than men overall, the distributions for both display a classic ``long-tailed'' form---most people receive a handful of messages at most, but a small fraction of the population receive far more.  The most popular individual in our four cities, a 30-year-old woman living in New York, received 1504 messages during the period of observation, equivalent to one message every 30 minutes, day and night, for the entire month.

However, desirability is not just about how many people contact you but also about who those people are.  If you are contacted by people who are themselves desirable, then you are presumptively more desirable yourself.  A standard measure of such reflected desirability is PageRank~\cite{BP98}.  Here we calculate PageRank scores for the populations within each of our four cities (\emph{Materials and Methods}, Network analysis), then rank men and women separately from least to most desirable.  A scaled rank of 1 denotes the most desirable man or woman in a city by our measure and 0 denotes the least desirable. It is important to emphasize that, while we use PageRank as an operational measure of desirability, we do not assume that users of the web site themselves use PageRank, or anything like it, to identify attractive mates.  In reality, a person might choose to message another based on an attractive profile picture, an interesting description, a good demographic match, an impressive income, or any of many other qualities.  PageRank scores simply give us, \emph{a posteriori,} a glimpse of who is desirable on aggregate, by identifying those people who receive the largest number of messages from desirable others.

\begin{figure*}
\begin{center}
\includegraphics[width=12cm]{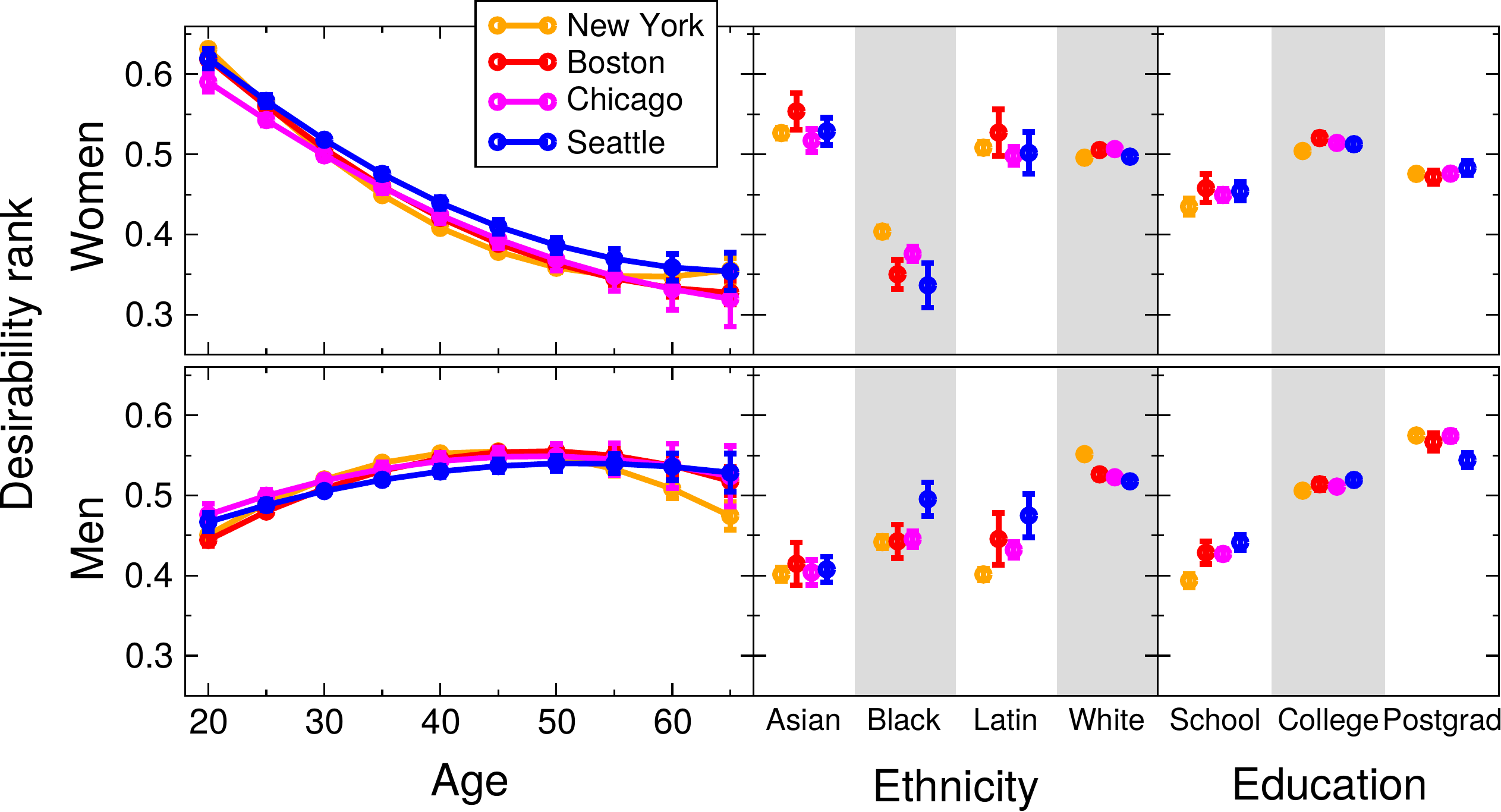}
\end{center}
\caption{Desirability, quantified using the measures defined here, as a function of demographic variables of the user population.  Leftmost panels show desirability as a function of age in years for women and men.  Middle panels show desirability by ethnicity.  Rightmost panels show desirability by highest educational level completed.  Error bars are $\pm1$ standard error. }
\label{fig:desirability}
\end{figure*}

Once we have our desirability scores, we can use them to identify characteristics of desirable users by comparing scores against various user attributes.  As shown in Fig.~\ref{fig:desirability}, for instance, average desirability varies with age for both men and women, though it varies more strongly for women and the effects run in opposite directions: older women are less desirable while older men are more so~\cite{rudder2015,england2009}.  For women this pattern holds over the full range of ages on the site: the average woman's desirability drops from the time she is 18 until she is 60.  For men desirability peaks around 50, then declines.  In keeping with prior work, there is also a clear and consistent dependence on ethnicity~\cite{lewis2013,Lin2013}, with Asian women and white men being the most desirable potential mates by our measures across all four cities.  The final panels in the figure show how desirability varies with educational level.  Desirability is associated with education most strongly for men, for whom more education is always more desirable. For women, an undergraduate degree is most desirable~\cite{HHA10}; postgraduate education is associated with decreased desirability among women.  These measurements control for age, so the latter observation is not a result of women with postgraduate degrees being older (\emph{Supplemental Information}, Table~\ref{tab:supplement2}).

\subsection{Reaching up the desirability ladder}

\begin{figure}
\begin{center}
\includegraphics[width=\columnwidth]{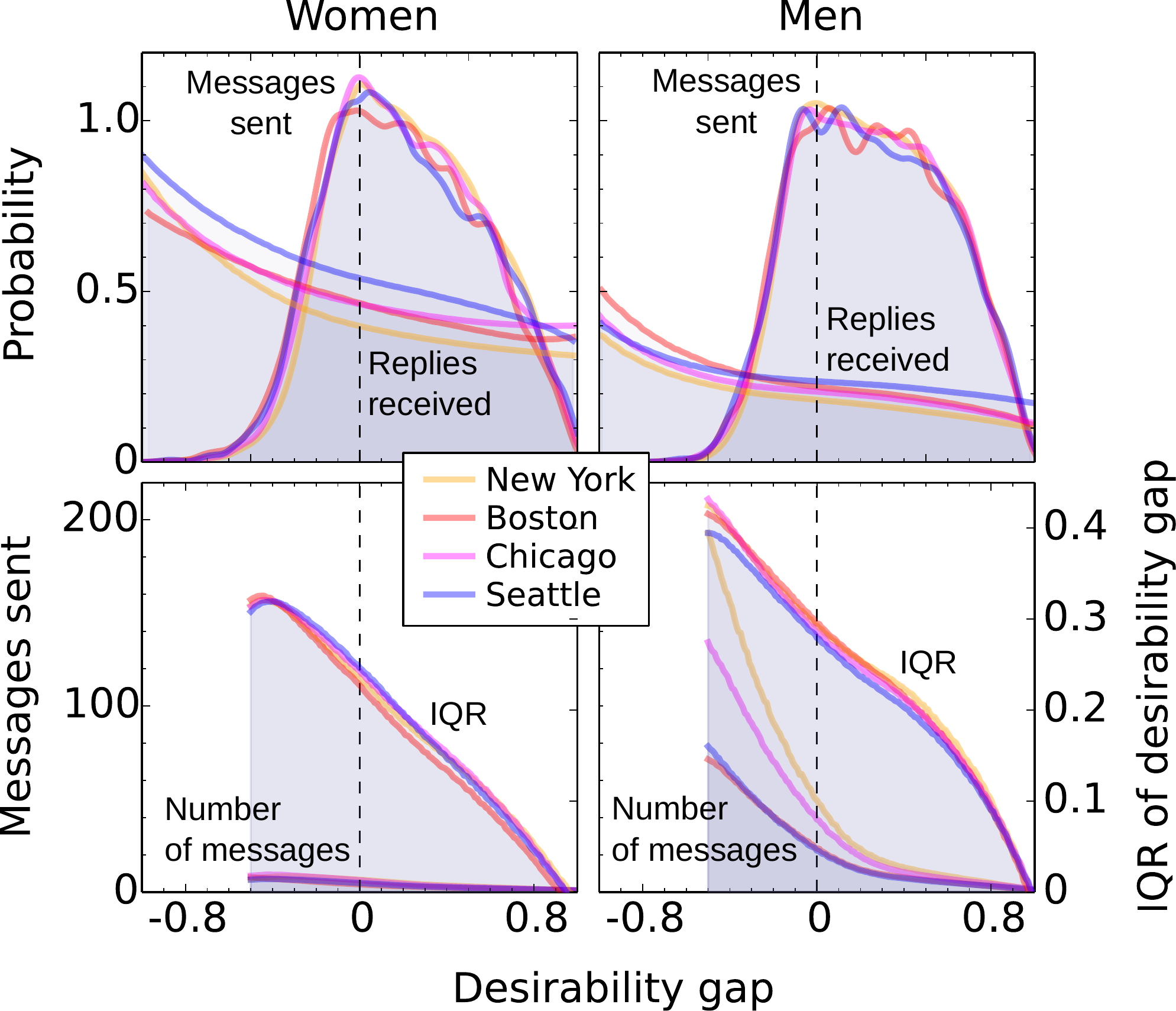}
\end{center}
\caption{Top panels: Upper curves show probability density for women and men of the median ``desirability gap,'' the difference in desirability rank of receiver and sender of an initial contact.  Both women and men tend to contact others who are ranked somewhat---but not excessively---higher than themselves.  The lower curves show the probability of receiving a reply to an initial message given the desirability gap between sender and receiver.  Women have higher overall probability of receiving replies, but both women and men have substantially lower probability of replies from more desirable partners.  Bottom panels: Lower curves show the average number of people contacted by individuals as a function of their average desirability gap.  Upper curves show the inter-quartile range of desirability of the people contacted, controlling for number of people contacted. Neither set of curves extends all the way to the left of the figure, because there is insufficient data to make reliable estimates in this regime.}
\label{fig:sent_rec_iqr}
\end{figure}

We now turn to the central results of our study.  First, we use our desirability scores to explore whether people engage in aspirational mate pursuit (i.e., messaging potential partners who are more desirable than they are) and how the probability of receiving a reply varies with the difference in desirability between senders and receivers.  In Fig.~\ref{fig:sent_rec_iqr} we show statistics for messages sent and replies received as a function of ``desirability gap,'' the difference in desirability ranking between the senders and receivers of messages.  If the least desirable man in a city were to send a message to the most desirable woman, the desirability gap would be~$+1$; if the most desirable man sent a message to the least desirable woman, the gap would be~$-1$.

The upper curves in the top panels of Fig.~\ref{fig:sent_rec_iqr} show the distribution of desirability gaps in our four cities.  For each individual we compute the median desirability gap over all initial messages they send and then plot the probability density of these numbers for men and women separately.  The most common (modal) behavior for both men and women is to contact members of the opposite sex who on average have roughly the same ranking as themselves, suggesting that people are relatively good judges of their own place in the desirability hierarchy.  The distributions about this modal value, however, are noticeably skewed to the right, meaning that a majority of both sexes tend to contact partners who are more desirable than themselves on average---and hardly any users contact partners who are significantly less desirable. The curves are remarkably consistent across all four cities, with men and women on average sending messages to potential partners who are 26\% and 23\% further up the rankings than themselves, respectively~\cite{note1}.

The lower set of curves in the top panels shows the probability of receiving a reply to an initial message.  The curves are higher for messages sent by women than for those sent by men---women are more likely than men to receive replies---but among both women and men the probability of a reply is a decreasing function of desirability gap, more desirable partners replying at lower rates than less desirable ones.  The differences are stark: men are more than twice as likely to receive a reply from women less desirable than themselves than from more desirable ones, and for messages sent to more desirable women the reply rate never rises above 21\%.  And yet the vast majority of men send messages to women who are more desirable than themselves on average.  Messaging potential partners who are more desirable than oneself is not just an occasional act of wishful thinking; it is the norm.

The bottom panels of Fig.~\ref{fig:sent_rec_iqr} show two further statistics that shed light on the mate seeking strategies adopted by users of the site.  The upper set of curves show the variation of desirability gaps across the potential partners a person contacts, quantified by inter-quartile range (the distance between the 25th and 75th percentiles in the distribution of desirability gaps).  Conditioned on the number of messages sent, men and especially women who reach higher up the desirability ladder tend to write to a less diverse set of potential matches, in terms of desirability gap. This behavior, consistent across all four cities, indicates that mate seekers, and particularly those setting their sights on the most desirable partners, do not adopt a diversified strategy to reduce the risk of being rejected, as one might for instance when applying to universities~\cite{bound2009}.

The lower set of curves in the bottom panels shows the average number of messages sent by a woman or a man as a function of average desirability gap.  Women initiate far fewer contacts than men, but both sets of curves fall off with increasing desirability gap in all four cities.  One might imagine that individuals who make a habit of contacting potential partners significantly more desirable than themselves (large positive desirability gap) would also initiate more contacts overall, to increase their chances of getting a reply, but in fact they do the opposite: the number of initial contacts an individual makes falls off rapidly with increasing gap and it is the people approaching the least desirable partners who send the largest number of messages.  A possible explanation is that those who approach more desirable partners are adopting a ``quality over quantity'' approach, more precisely identifying people they see as an attractive match or spending more time writing personalized messages, at the expense of a smaller number of messages sent.

\begin{figure}
\begin{center}
\includegraphics[width=\columnwidth]{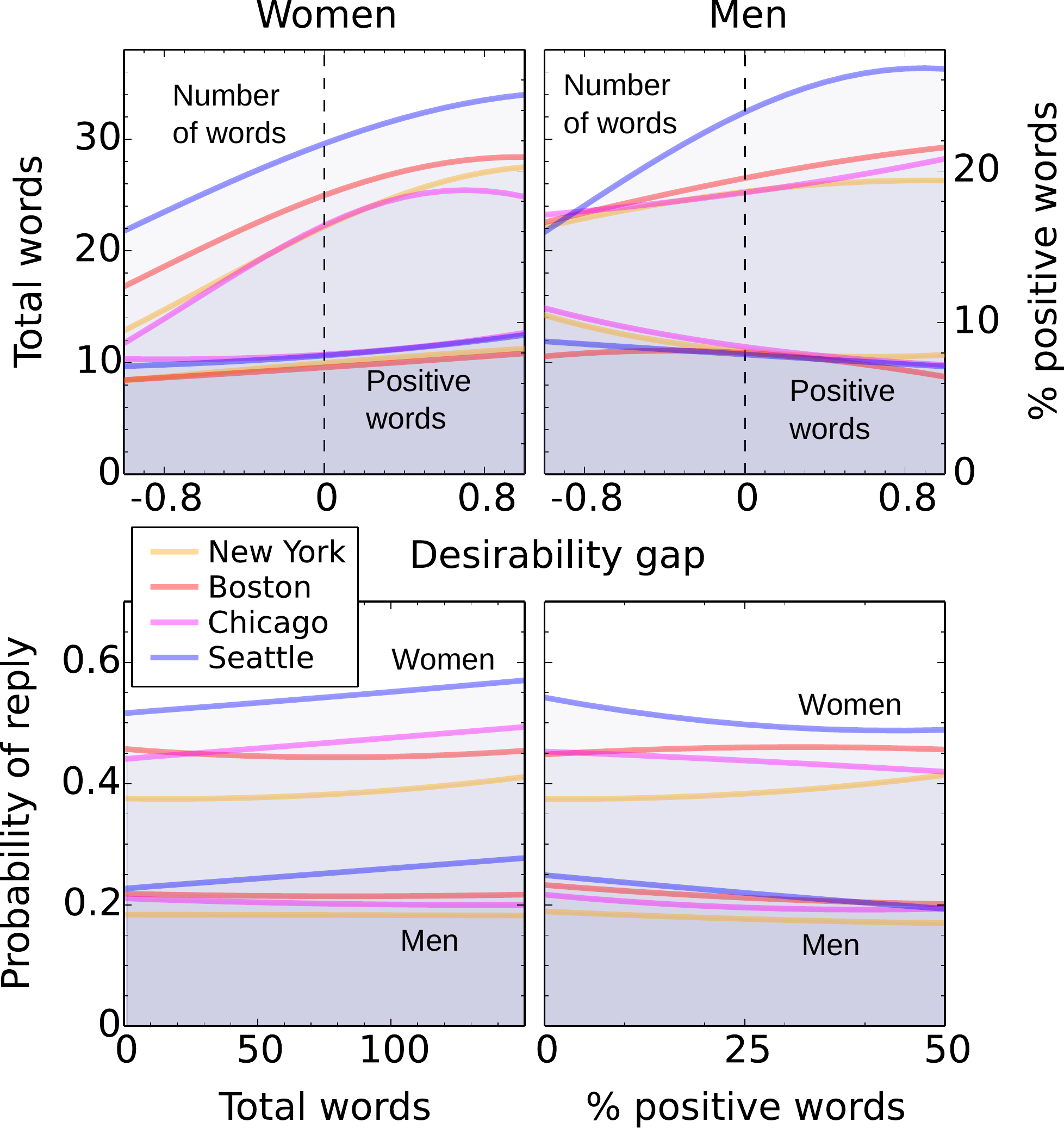}
\end{center}
\caption{Top panels: Upper curves show the total number of words in initial messages, which increases with desirability gap.  Lower curves show the fraction of positive words in messages, which increases slightly for messages sent by women but decreases for messages sent by men. Bottom panels: Expected payoffs to writing longer and more positive messages, holding desirability gap at its city-specific mean. We see that longer messages are positively associated with response rates only for women and for men in Seattle. Positive messages are somewhat negatively associated with response rates for men; women have mixed success with more positive messages, depending on the city.}
\label{fig:strategy}
\end{figure}

\subsection{Messaging strategies vary with mate seekers' aspirations}
Do mate seekers put more effort into attracting more desirable partners?  Based on message content, there is some evidence that they do.  In the top two panels of Fig.~\ref{fig:strategy} the upper set of curves show how the total length in words of initial messages sent varies by desirability gap.  Both men and women tend to write substantially longer messages to more desirable partners, up to twice as long in some cases.  The effect is larger for messages sent by women than by men, though there are exceptions.  Among the groups we study, for instance, it is men in Seattle who have the most pronounced increase in message length (see also \emph{Supplemental Information}, Table~\ref{tab:supplement3})~\cite{note2}.

The lower set of curves in the same panels shows a simple measure of the emotional content of messages, the fraction of positive words (based on the LIWC2001 database~\cite{PFB01,kahn2007}).  Here we see an interesting difference between women and men, the women showing an increase in their use of positive words when communicating with more desirable partners, while men show a decrease.  The effect size is modest but consistent across all four cities and statistically significant ($p<0.001$, \emph{Supplemental Information}, Table~\ref{tab:supplement4}).

The bottom panels of Fig.~\ref{fig:strategy} quantify the payoffs to writing longer or more positive messages, controlling for the desirability gap between senders and receivers (\emph{Supplemental Information}, Section~\ref{sec:replyprob}). The expected payoffs for both men and women show a remarkably close match to the messaging behavior depicted in the upper panels. For example, in all four cities men experience slightly lower reply rates when they write more positively worded messages. Though our analysis cannot reveal the underlying process that gives rise to these behaviors (e.g., reinforcement learning), this result may offer a hint about why men tend to write somewhat less positive messages to more desirable partners~\cite{note3}.  Similarly, only Seattle men experience a payoff to writing longer messages---and Seattle is the only city where men write longer messages to more desirable mates. Overall, however, the variation in payoff for different strategies is fairly small, suggesting that, all else being equal, effort put into writing longer or more positive messages may be wasted.

\section{Discussion}

The results presented here provide a picture of the aspirational pursuit of mates in online dating and its implications for the likelihood of success. We present a network measure of desirability in dating that is based on mate seeking behavior rather than subjective personal qualities like attractiveness. We find that while some mate seekers do pursue partners of similar average desirability to themselves, the vast majority of the online dating population we study tend to reach up the hierarchy toward more desirable partners.  At the same time, this aspirational mate pursuit is calibrated to one's own desirability: on average, people pursue partners who are roughly 25\% more desirable than they themselves are.  In the language of matching and competition introduced at the start of this article people are, it appears, pursuing a hybrid strategy with elements of both---they are aware of their own position in the hierarchy and adjust their behavior accordingly, while at the same time competing modestly for more desirable mates. 

We find that all but the most extreme mate seekers exhibit heterogeneity in their mate pursuit, initiating contact with partners across a range of desirabilities.  This suggests that both men and women combine aspirational mate pursuit with less risky prospects.  Additionally, there appears to be a ``quality over quantity'' strategy such that men and women who pursue more desirable partners send fewer messages, each with a higher word count on average.  Message strategies also become less diversified (in terms of range of desirability gaps) as people reach higher up the desirability ladder.

Our results on aspirational mate pursuit are consistent with the popular concept of dating ``leagues,'' as reflected in the idea that someone can be ``out of your league,'' meaning that attractive matches are desirable for but unavailable to less attractive others.  Provided leagues are envisaged as a single continuous hierarchy rather than as distinct strata, our results suggest that, contrary to popular belief, attracting the attention of someone out of one's league is entirely possible. The chances of receiving a reply from a highly desirable partner may be low but they remain well above zero, although one will have to work harder, and perhaps also wait longer~\cite{Kalick1986}, to make progress.  Compared to the extraordinary effort male rats are willing to go through to mate with a desirable female~\cite{Everitt1990}, however, messaging 2 or 3 times as many potential partners to get a date seems quite a modest investment.

One might wonder how the patterns we observe online might inform our understanding of offline mate pursuit and dating markets. Online dating differs from offline dating in several important ways~\cite{finkel2012}.  Because of the high volume of partners and low threshold for sending a message, competition for potential partners' attention is likely fiercer online than offline. This may increase the extent to which there exists a hierarchy of desirability online, and reduce people's willingness to respond to less desirable mates: when there are plenty of fish in the sea one can afford to throw a few back.  It has also been suggested that consensus about what makes an attractive partner is strongest in the early stages of courtship, when partners do not know as much about one another~\cite{eastwick2014,hunt2015}. While it is difficult to study early courtship offline---our method requires unrequited overtures, which are hard to observe in offline interactions---these differences suggest that hierarchies of desirability may be more pronounced online than off.

\section{Materials and Methods}

\subsection{Data}
Online dating has grown greatly in popularity in recent years and has become an increasingly common way for people to find romantic partners, edging out more traditional means such as meeting through coworkers or through family. By 2013, the Pew Research Center~\cite{smith2013} found that 11\% of all American adults, and 38\% of those who were currently single and searching for a partner, had used online dating sites or mobile apps.  Two-thirds of online daters had gone on a date with someone they met through a site and almost a quarter (23\%) had entered into a marriage or long-term relationship with someone they met through a site.  Thus, online dating now plays a substantial role in the organization of sexual and romantic relationships in the United States---it is currently the third most common way partners meet after meeting through friends or in bars~\cite{rosenfeld2011,note4}.

The data used as the starting point for our study consist of demographics and messaging patterns for active users of a popular online dating site during a one month period of observation from January 1 to 31, 2014.  The site does not market itself to any particular demographic group and attracts a diverse population of users whose makeup, in most locales, corresponds loosely to that of the general population. The population of users is concentrated in coastal areas, although there are significant numbers of users in major midwestern cities such as Chicago. Upon joining the site, users specify a login handle and enter their age, sexual orientation, relationship status, and a 5-digit ZIP code identifying their location.  All but the ZIP code are visible to other users, while geographic location is publicly listed at the city level.  Optionally, users can also give additional demographic information (e.g.,~height, religion, body type) and answer a set of open-ended essay questions that ask them to describe who they are and what they are looking for.  After creating a profile, users can then view the profiles of others, as well as send and receive messages.

In addition to demographics our data include complete messaging patterns---who sends messages to whom on the site.  It is these messages that we use to assess individuals' desirability.  We restrict our analysis to active users, which we define to mean users who sent or received at least one message during the observation period.  This eliminates a significant number of users who sign up and use the site but then become inactive, or who sign up and never use it.  For the purposes of the present study we also remove from the data all users who identify as gay or bisexual (about 14\% of the overall user base of the site) and those who indicate that they are not looking for romantic relationships.  (People can indicate, for example, that they are only looking for friendship or activity partners.)  Details about the demographic makeup of users in each city are shown in Section~\ref{sec:statistics} of the \emph{Supplemental Information}.

We report results for four large metropolitan areas---New York City, Boston, Chicago, and Seattle.  One reason for restricting our study to individual cities is to reduce the effects of spatial distance in mate selection behavior: we choose areas large enough to give good demographic statistics but small enough geographically that distance will not be a significant deterrent to conversation between interested users.  In the case of Boston, Chicago, and Seattle, we find a good choice to be the standard Core Based Statistical Areas (CBSAs) established by the Office of Management and Budget.  A CBSA is defined to be an urban center of at least $10\,000$ people plus adjacent areas that are socioeconomically tied to the urban center by commuting.  For New York City the standard CBSA proves too large: the data clearly indicate multiple geographic dating markets within the larger metro area.  Instead, therefore, we choose a narrower set of geographic boundaries for New York, the five boroughs of Manhattan, the Bronx, Queens, Brooklyn, and Staten Island. Some descriptive statistics for the user populations in the four cities are reported in the \emph{Supplemental Information}, Table~\ref{tab:data}. Restricting our study to metropolitan areas inevitably eliminates some messaging activity to and from outlying regions, but the areas chosen here capture a large majority of the messaging activity of the users who live in them.

\subsection{Network analysis}

We construct a network for each city studied in which the nodes represent users, and connections between nodes---directed edges in network nomenclature---represent the first message sent in the corresponding direction between any two users.  That is, there is a directed edge in the direction of the initial contact between two users and, optionally, a second edge in the opposite direction if that initial contact received a reply.  Our analyses are based on the largest weakly connected component of the network in each city, although in practice this restriction has little effect since nearly everyone belongs to the largest component.  In the network for New York City, for example, the largest weakly connected component contains 99.8\% of all users.

Given that our focus here is on who is interested in whom, one approach might be to restrict ourselves to a network with edges representing only the first direction of contact between individuals and excluding any reply.  A defining feature of heterosexual online dating, however, is that in the vast majority of cases it is men who establish first contact---over 80\% of first messages are from men in our data set.  As a result there is little information about women's aspirations contained in first messages.  On the other hand, women reply very selectively to the messages they receive from men---their average reply rate is less than 20\%---so women's replies (along with the small fraction of first messages sent by women) can give us significant insight about who they are interested in.  To create a picture of both men's and women's aspirations, therefore, we include both first messages and replies in our network.

A related challenge is how to choose which users should be included in the network.  One approach might be to restrict our list of active users to those who sent at least one message during the observation period.  However, because, again, men send most messages, this would exclude a large number of women from the sample.  To avoid this we choose to include in our networks all users who either sent or received at least one message during the period of observation.

\subsubsection{Desirability rankings}
The directed network of initial contacts is used as the starting point for our PageRank-based measure of desirability.  In this calculation, network nodes are first numbered, in arbitrary order, from 1 to $n$, where $n$ is the total number of nodes in the network, and then we assign each node~$i$ (i.e.,~each person) a positive desirability score~$x_i$.  The structure of the network itself is represented by the directed adjacency matrix~$A$ having elements~$a_{ij}=1$ if there is a directed edge from node~$j$ to node~$i$ and zero otherwise.  Then the scores obey the standard PageRank equation~\cite{BP98}
\begin{equation}
x_i = 1 + \alpha \sum_{j=1}^n {a_{ij} x_j\over\sum_{k=1}^n a_{kj}},
\label{eq:pagerank}
\end{equation}
where $\alpha$ is a parameter whose value we choose.  There is no formal theory specifying the best value of this parameter, but the inventors of the PageRank method~\cite{BP98} recommend a value of $\alpha=0.85$ and we use that value here.  (Our results are not particularly sensitive to the value of~$\alpha$---calculations with other values lead to qualitatively similar conclusions.)  The numerical solution of equation~\eqref{eq:pagerank} is straightforward: one starts with any set of nonnegative values~$x_i$, such as $x_i=0$ for example, and uses them to evaluate the right-hand side of~\eqref{eq:pagerank}, giving a new set of values~$x_i'$.  Then one substitutes these into the equation again to calculate another new set, and repeats the process until the values converge within a desired accuracy.  For networks of the size studied here the calculation takes less than a second on a standard desktop computer.

There is an extensive literature on network measures of social rank. However, only a small handful of studies have used network measures to explore how social rank is associated with mating success~\cite{Ryder2009, Page2017,Place2010}.  These studies all employ eigenvector centrality, a matrix-based measure similar in some respects to PageRank but designed for use with undirected networks. These studies have focused primarily on small populations (two hunter-gatherer societies, leks of birds, men and women in a speed dating experiment). Our study is the first we know of to apply PageRank scores as a measure of desirability in large-scale online dating populations. 

Further details about the statistical models used in the analysis, as well as estimated coefficients, can be found in the \emph{Supplementary Information}. 

\begin{acknowledgments}
This work was funded in part by the National Institutes of Health under grant K01--HD--079554 (EEB) and the National Science Foundation under grants DMS--1107796 and DMS--1407207 (MEJN).
\end{acknowledgments}

\section*{Supplementary materials}
\renewcommand{\theequation}{S\arabic{equation}}
\setcounter{equation}{0}
\renewcommand{\thefigure}{S\arabic{figure}}
\setcounter{figure}{0}
\setcounter{section}{0}

\section{Background literature}
In the introduction, we presented two competing hypotheses of mate preferences: matching and competition.  Here we expand on this discussion and provide an evaluation of previous efforts to study desirability in dating markets.

A variant of the matching hypothesis, originally proposed by Walster~\cite{walster1966} and now considered a classic in the psychology literature, posits that all (or most) men and women do find the same attributes desirable and are attracted to the same potential partners, and they might as a result feel the urge to pursue someone ``out of their league,'' as the competition hypothesis would suggest.  They do not, however, either for fear of rejection, or simply to maximize their chances of success. Instead, they take into account their own desirability when deciding who to pursue~\cite{todd1999,todd2007,penke2007,fletcher2013,walster1966,walster1971} and only approach mates of desirability similar to their own~\cite{kirsner2003,kavanagh2010}.  The net result, once again, is an assortative pairing of like with like.

Walster's work and subsequent studies building upon it assume that men and women have an overall desirability or ``mate value,'' a score by which people, implicitly or explicitly, rank themselves and others.  Our work has something in common with this approach in that, rather than relying on personal characteristics, we quantify desirability using empirical measures of who actually receives the most attention and from whom.  This lends itself to the study of hierarchy in dating markets---our primary focus in this paper---by placing all actors on a single scale and allowing us to quantify concepts such as reach and desirability gap.

If one were to observe matching according to such overall desirability measures it would suggest either that people prefer partners of similar desirability or, as Walster hypothesizes, pursue partners of similar desirability for pragmatic reasons.  In practice this is observed to some extent---most people pursue partners whose desirability is not too different from their own---but at the same time most people are ``aiming high,'' with the average potential partner pursued having somewhat higher desirability than their pursuer.  It is in this sense that we describe the observed behavior as a hybrid of matching and competition behaviors.

\subsection{Previous work on desirability and matching}
Previous studies have not provided a crisp formal definition of desirability or articulated a clear link between desirability and mate-seeking behavior.  Related concepts have, however, arisen, particularly in the work of Fiore, Taylor, and coauthors~\cite{fiore2010,taylor2011}.  It is useful to discuss their work in some detail, since they also make use of large-scale online dating data.

The most relevant study is that of Taylor~\etal~\cite{taylor2011}, in which the authors consider empirical evidence for Walster's matching hypothesis. They define a desirability score based on popularity, as measured by the number of first messages received on a dating site, and report a number of findings.  First, they find that there is a weak but positive correlation between the popularity of message senders and receivers ($R$-values of 0.27 for messages sent by women and 0.37 for messages sent by men). They interpret this as evidence of popularity-based matching.  Second, they assess whether people who match in this way are more successful, in terms of receiving replies, than those who do not.  They compute the mean absolute difference between users' own popularity and the popularity of people they contacted.  Their motivation for looking at absolute differences is that the matching hypothesis predicts that it should be equally disadvantageous to contact people who are either more or less desirable than yourself.  They correlate each user's average success rate with their average absolute popularity gap, and find that this correlation is very close to zero ($R$-values of $-0.01$ for men and $0.00$ for women).  Taylor~\etal\ interpret these results as indicating that both men and women tend to contact partners who are ``in their league'', but that their chances of getting a reply would not be affected if they did not.

This interpretation is in stark contrast to our own findings.  We find that both men and women tend to message up the desirability hierarchy, and that there is a pronounced drop in the probability of reply with increasingly positive desirability gap.  There are a number of methodological differences between our approach and that of Taylor~\etal\ that could account for this disagreement.  To start with, Taylor~\etal\ use simple popularity---number of messages received---as a proxy for desirability, which immediately introduces difficulties: if more desirable individuals receive more messages it implies that either (a)~they are receiving a lot of messages from individuals less desirable than themselves, which would run counter to Taylor~\etal's claims that this is not happening, or (b)~they are receiving messages from people of similar high desirability but such high-desirability people are sending more messages on average than low desirability people, which is the opposite of what we observe to be true.

If we nonetheless accept popularity as a measure of mate desirability, the results reported by Taylor~\etal\ are not sufficient to prove the presence of matching in their data for two reasons.  First, the existence of a correlation between the popularity of message senders and receivers, even a much stronger one than the authors find, cannot be used as evidence for matching, since correlation $R$-values of the type used by Taylor~\etal\ are unaffected by uniform differences in popularity.  In other words, one could achieve the same $R$-value if all individuals were messaging others more popular than themselves as one would if they were messaging others of the same popularity.  The $R$-value is simply not sensitive to the absolute value of popularity and hence cannot be used as a measure of matching~\cite{note5}.

Second, the authors' analysis of the relationship between desirability gap and reply probabilities conceals from view much of the dependence of replies on receivers' desirability.  Recall that the authors measure the correlation between users' average absolute desirability gap and average rate of receiving replies.  Use of the average gap obscures any variation of reply rate with desirability gap for an individual user, and, more importantly, the focus on \emph{absolute} gap size means that the analysis cannot distinguish effects of messaging up versus down the desirability hierarchy.  Our study reveals that people who message down the desirability hierarchy have a higher chance of reply than people who seek out partners of similar desirability, and people who message up the hierarchy have a lower chance of reply than people who seek out partners of similar desirability.  Taylor~\etal\ by contrast conclude that there is no effect, but this appears to be an artifact of the way their analysis combines positive and negative gaps in a single measure, resulting in an average change in reply rate close to zero.

Our finding, based on the PageRank measure of desirability, is that individuals are in fact \emph{not} matching on desirability.  It is true that there is a positive correlation between desirability of sender and receiver, and most users message others of desirability not too dissimilar from their own, but there is also an offset, with most focusing on potential partners of higher rather than lower desirability.

In this sense our findings are different from those of Taylor~\etal, and yet we still conclude that there is a hierarchy of desirability within the community we study and that the patterns of who pursues whom are strongly correlated with it.  How can we make this statement in the absence of strong matching?  The crucial observation is that most users message across a range of desirabilities, but receive replies only when they send messages to others who are of similar or lower desirability to themselves.  Thus the observed behavior seems to be a hybrid of the traditional matching and competition models.  On the one hand, people appear to be aware of their place in the desirability hierarchy and make their overtures accordingly, since on average they send messages to others who are not greatly more desirable than themselves.  One could say therefore that a weak (and biased) form of matching is taking place.

\begin{table*}
\label{table1}
\centering
\begin{tabular}{lcccccccc}
\multicolumn{1}{c}{} &  \multicolumn{2}{c}{New York} & \multicolumn{2}{c}{Boston} & \multicolumn{2}{c}{Chicago} & \multicolumn{2}{c}{Seattle} \\
                       & Men           & Women         & Men           & Women         & Men           & Women         & Men           & Women         \\
\hline
Total number of users  & $44\,009$     & $50\,618$     & $9\,113$      & $9\,355$      & $28\,635$     & $23\,236$     & $12\,721$     & $9\,248$      \\
Ethnicity (\%)         &               &               &               &               &               &               &               &               \\
 \quad Asian          &       8       &       11      &       4       &       6       &       3       &       4       &       7       &       9       \\
 \quad Black          &       9       &       9       &       6       &       6       &       7       &       9       &       4       &       3       \\
 \quad Hispanic       &       10      &       8       &       3       &       3       &       8       &       7       &       3       &       3       \\
 \quad White          &       73      &       73      &       87      &       85      &       81      &       80      &       87      &       85      \\
College degree (\%)    &       92      &       96      &       70      &       80      &       63      &       71      &       64      &       68      \\
Children at home (\%)  &       5       &       6       &       7       &       10      &       7       &       10      &       15      &       17      \\
Mean age               &       31.6    &       31.5    &       30.4    &       30.3    &       31.4    &       32      &       32.7    &       33.1    \\
Mean messages sent     &       23.3    &       9.4     &       14.6    &       6.3     &       19      &       10.2    &       12.4    &       7.8     \\
Replies received (\%) & 15    &       34      &       17      &       37      &       18      &       40      &       20      &       45      \\
\hline
\end{tabular}
\caption{User attributes for four metropolitan areas.\label{tab:data}}

\end{table*}

This critique also highlights the danger of conflating matching with the existence of a hierarchy of desirability.  While the matching hypothesis---as Walster formulated it---is related to hierarchy, the two do not necessarily imply the same behaviors.  Most notably, the matching hypothesis implies that men and women do not engage in aspirational mate pursuit.

While our analysis is largely descriptive in nature, our hope is that such rich description can improve social scientists' theoretical understanding of the behaviors that produce hierarchies in dating markets.  Large-scale activity datasets such as those produced through online dating provide a unique opportunity to study human behavior at a high level of granularity.  However, such data require our theories and models to move beyond the conceptual architecture developed for analyses of surveys and administrative data.

\section{Descriptive statistics}
\label{sec:statistics}
\begin{figure}
\centering
\includegraphics[width=\columnwidth]{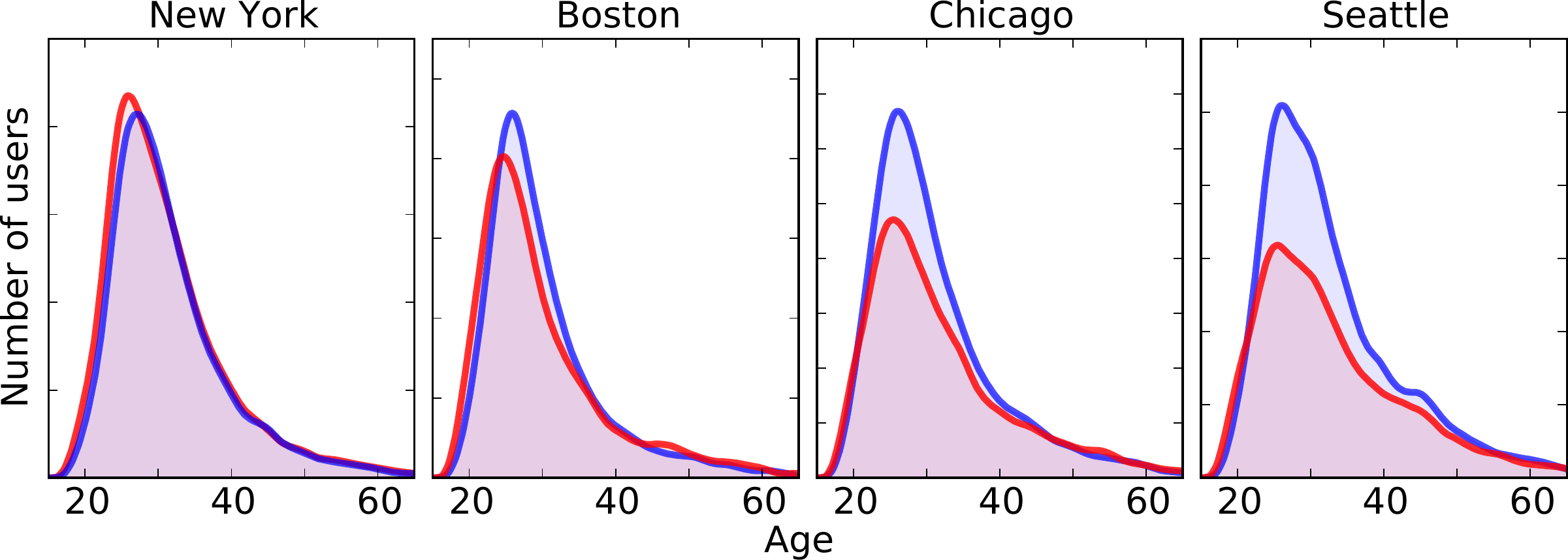}
\caption{Age distribution of men (blue) and women (red) in each city.  Seattle and Chicago, and to a lesser extent Boston, have surpluses of men, the surplus being most pronounced for people around 28 years of age.  New York city has a surplus of women, which is most pronounced among women in their mid-twenties.  Note that because the total number of users varies across cities, the scale of the $y$-axis differs across the four panels.}
\label{fig:age}
\end{figure}

Table~\ref{tab:data} gives a set of summary statistics for the male and female user populations in each of the four cities that are the focus of our study.  The cities display a range of values of the ratio of men to women, New York having the largest fraction of women, followed by Boston, Chicago, and Seattle, in that order. Overall, the site has approximately 55 men for every 45 women.  This slight excess of men is consistent with other studies of US online dating~\cite{HHA10,Lin2013,lewis2013}.  In addition to their sex ratios, the cities differ in their overall market size and composition.  New York City has the largest number of users, followed by Chicago, Seattle, and Boston.  The average user is in their early 30s in all cities but there is modest variation in this and other demographic characteristics.  Seattle users, for example, are slightly older and are more likely to have children living at home. Figure~\ref{fig:age} shows the age distribution of men and women in each city. We see that New York has a surplus of women which is most pronounced among users in their mid twenties.  The remaining cities all have a surplus of men, which is most pronounced in the later 20s and early~30s.

Table~\ref{tab:data} also shows the average number of initial contacts made by men and women in each city and the percentage of those contacts that receive replies.  As observed in other studies~\cite{HHA10,Lin2013,lewis2013}, men send significantly more messages than women.  Overall they are responsible for 81\% of initial contacts on the site, but men have a lower chance than women of receiving replies to their messages.  This is not surprising: women may well reply less often precisely because they receive so many messages.  The number of messages sent does, however, show some interesting variation between cities.  Notice, for example, that among the cities studied men send the smallest number of messages and experience the largest reply rate in Seattle, which is unexpected since this is the poorest dating environment for men in terms of ratio of men to women.  One might imagine that in cities where the sex ratio puts men at a disadvantage they would send more messages, in the hope of getting a reasonable number of replies.  Moreover, the low number of messages sent by men in Seattle cannot be explained as a result of a larger fraction of inactive users, which might occur if male users become discouraged by the poor dating environment.  As described above, only active users are included in the data, although it is possible that Seattle might contain a larger-than-usual number of users of low (but nonzero) activity level.

\section{Supplementary Analyses}
In this section we describe the statistical models used to create Figs.~\ref{fig:desirability} and~\ref{fig:strategy} in the main paper.

\subsection{Desirability as a function of selected demographics}
Figures for average desirability as a function of demographic characteristics shown in Fig.~\ref{fig:desirability} were calculated using fractional regression models of men and women's relative desirability as a function of their attributes.  We specify robust standard errors to allow for clustering of observations within cities, while model interactions allow the effects of demographics on relative desirability to vary by city.  The main effects refer to the values for Boston.  Coefficient estimates from the fractional regression models are shown in Table~\ref{tab:supplement2}.  In addition to those covariates included in the table, we control for other user attributes that might affect desirability: users' body type, whether they have children, and the type of relationship sought (e.g.,~short-term relationship, long-term relationship, or sex). A complete set of coefficients are available from the authors by request.

\subsection{Message length and positivity by desirability gap}
The top panels of Fig.~\ref{fig:strategy} show the relationship between desirability gap and message attributes.  The predicted values of message length are derived from negative binomial regressions where the outcome is the total word count of the first message and the predictor variables are linear and quadratic terms for desirability gap. The predicted values of message positivity are derived from a fractional regression model where the outcome is the proportion of words in the message that are positively valenced~\cite{PFB01,kahn2007}, and the predictor variables are linear and quadratic terms for desirability gap. Separate effects are estimated for each city via dummy variable interactions.  The complete set of coefficients are shown in Tables~\ref{tab:supplement3} and~\ref{tab:supplement4}.  The units of observation are first messages sent by a particular mate seeker to a potential match.  Standard errors are robust to allow for clustering within mate seekers.  To aid in ease of interpretation and presentation of results without excessive significant digits, the number of words in messages is divided by~100. (The values are scaled back up to their original levels in Fig.~\ref{fig:strategy}.)

\subsection{Reply probabilities by message length and positivity}
\label{sec:replyprob}
The bottom panels of Fig.~\ref{fig:strategy} show expected reply rates as a function of message length and positivity.  These values are derived from logistic regressions describing how the probability of receiving a reply to an initial contact varies with message length and the percent of words in the message that have positive connotations~\cite{PFB01,kahn2007}.  Because online dating site users, especially women, tend to write longer and more positive messages to more desirable partners, we control for the desirability gap between sender and receiver.  We also allow for potentially nonlinear effects of desirability, message length, and message positivity on the probability of receiving a reply.  Separate effects are estimated for each city via dummy variable interactions.  The complete set of coefficients are shown in Tables~\ref{tab:supplement5} and~\ref{tab:supplement6}.  The units of observation are first messages sent by a particular mate seeker to a potential match.  Standard errors are robust to allow for clustering within mate seekers.  To aid in ease of interpretation and presentation of results without excessive significant digits, the number of words in messages is divided by~100. (The values are scaled back up to their original levels in the article figures.)

\begin{table*}
\begin{center}
\begin{tabular}{lrrrrrr}
\multicolumn{1}{c}{} &  \multicolumn{3}{c}{Women} & \multicolumn{3}{c}{Men} \\
 & Coef. & Std. error & $z$ & Coef. & Std. error & $z$ \\
 & \multicolumn{3}{c}{\hrulefill} & \multicolumn{3}{c}{\hrulefill} \\
Age & $-0.055$ & $0.014$ & $-3.840$ & $0.036$ & $0.014$ & $2.480$ \\
Chicago $\times$ Age & $-0.019$ & $0.016$ & $-1.160$ & $0.019$ & $0.016$ & $1.160$ \\
NYC $\times$ Age & $-0.040$ & $0.016$ & $-2.530$ & $0.030$ & $0.016$ & $1.860$ \\
Seattle $\times$ Age & $-0.012$ & $0.019$ & $-0.660$ & $-0.006$ & $0.018$ & $-0.310$ \\
 \\
Age$^2$ & $0.000$ & $0.000$ & $1.860$ & $0.000$ & $0.000$ & $-1.990$ \\
Chicago $\times$ Age$^2$ & $0.000$ & $0.000$ & $0.940$ & $0.000$ & $0.000$ & $-0.930$ \\
NYC $\times$ Age$^2$ & $0.000$ & $0.000$ & $2.260$ & $0.000$ & $0.000$ & $-1.830$ \\
Seattle $\times$ Age$^2$ & $0.000$ & $0.000$ & $0.630$ & $0.000$ & $0.000$ & $0.350$ \\
 \\
Black & $-0.729$ & $0.126$ & $-5.770$ & $0.227$ & $0.142$ & $1.600$ \\
Latino/a & $-0.014$ & $0.153$ & $-0.090$ & $0.278$ & $0.173$ & $1.600$ \\
White & $-0.110$ & $0.099$ & $-1.120$ & $0.492$ & $0.114$ & $4.330$ \\
 \\
Black $\times$ Chicago & $0.301$ & $0.146$ & $2.070$ & $0.057$ & $0.162$ & $0.350$ \\
Black $\times$ NYC & $0.275$ & $0.132$ & $2.080$ & $0.038$ & $0.151$ & $0.250$ \\
Black $\times$ Seattle & $0.124$ & $0.195$ & $0.640$ & $0.173$ & $0.179$ & $0.970$ \\
Latin $\times$ Chicago & $0.015$ & $0.171$ & $0.090$ & $0.000$ & $0.190$ & $0.000$ \\
Latin $\times$ NYC & $0.004$ & $0.159$ & $0.030$ & $-0.112$ & $0.181$ & $-0.620$ \\
Latin $\times$ Seattle & $0.047$ & $0.201$ & $0.230$ & $0.108$ & $0.217$ & $0.500$ \\
White $\times$ Chicago & $0.182$ & $0.116$ & $1.570$ & $0.045$ & $0.132$ & $0.340$ \\
White $\times$ NYC & $0.014$ & $0.103$ & $0.130$ & $0.101$ & $0.120$ & $0.840$ \\
White $\times$ Seattle & $0.147$ & $0.123$ & $1.190$ & $-0.019$ & $0.133$ & $-0.150$ \\
 \\
No college & $-0.082$ & $0.081$ & $-1.010$ & $-0.305$ & $0.066$ & $-4.590$ \\
Post-college & $-0.132$ & $0.045$ & $-2.900$ & $0.174$ & $0.053$ & $3.290$ \\
No college $\times$ Chicago & $-0.050$ & $0.090$ & $-0.560$ & $0.025$ & $0.073$ & $0.340$ \\
No college $\times$ NYC & $-0.039$ & $0.094$ & $-0.420$ & $0.005$ & $0.077$ & $0.060$ \\
No college $\times$ Seattle & $-0.060$ & $0.100$ & $-0.600$ & $0.009$ & $0.080$ & $0.110$ \\
Post-college $\times$ Chicago & $0.090$ & $0.053$ & $1.710$ & $0.006$ & $0.061$ & $0.110$ \\
Post-college $\times$ NYC  & $0.130$ & $0.049$ & $2.660$ & $0.006$ & $0.057$ & $0.110$ \\
Post-college $\times$ Seattle & $0.112$ & $0.064$ & $1.760$ & $-0.122$ & $0.069$ & $-1.770$ \\
 \\
\multicolumn{1}{l}{$N$} & & $32\,832$ & & & $31\,725$ \\
\multicolumn{1}{l}{Log-likelihood} & & $-22\,247$ & & & $-21\,662$ \\
\end{tabular}
\caption{Fractional regression of desirability on individual attributes (selected coefficients).\label{tab:supplement2}}
\end{center}
\end{table*}

\begin{table*}
\begin{center}
\begin{tabular}{lrrrrrrr}
 & & \multicolumn{3}{c}{Women} & \multicolumn{3}{c}{Men} \\
 & & Coef. & Std. error & $z$ & Coef. & Std. error & $z$ \\
 & & \multicolumn{3}{c}{\hrulefill} & \multicolumn{3}{c}{\hrulefill} \\
constant &  &  $3.218$	& $0.012$	& $268.1$ &  $3.277$ & $0.005$ & $681.5$ \\
Chicago &  & $-0.114$ & $0.013$ & $-8.600$ & $-0.050$ & $0.005$ & $-9.560$ \\
NYC &  & $-0.120$ & $0.013$ & $-9.520$ & $-0.047$ & $0.005$ & $-9.280$ \\
Seattle &  & $0.170$ & $0.016$ & $10.810$ & $0.201$ & $0.006$ & $32.140$ \\
 \\
Desirability Gap &  & $0.262$ & $0.025$ & $10.390$ & $0.130$ & $0.010$ & $12.740$ \\
Desirability Gap $\times$ Chicago &  & $0.113$ & $0.028$ & $4.010$ & $-0.033$ & $0.011$ & $-2.930$ \\
Desirability Gap $\times$ NYC &  & $0.118$ & $0.027$ & $4.390$ & $-0.045$ & $0.011$ & $-4.140$ \\
Desirability Gap $\times$ Seattle &  & $-0.040$ & $0.033$ & $-1.210$ & $0.127$ & $0.013$ & $9.570$ \\
 \\
Desirability Gap$^2$ &  & $-0.134$ & $0.050$ & $-2.670$ & $-0.031$ & $0.020$ & $-1.570$ \\
Desirability Gap$^2$ $\times$ Chicago &  & $-0.133$ & $0.056$ & $-2.380$ & $0.047$ & $0.022$ & $2.150$ \\
Desirability Gap$^2$ $\times$ NYC &  & $-0.030$ & $0.053$ & $-0.570$ & $-0.016$ & $0.021$ & $-0.730$ \\
Desirability Gap$^2$ $\times$ Seattle &  & $0.049$ & $0.066$ & $0.750$ & $-0.113$ & $0.026$ & $-4.350$ \\
 \\
$N$  &  &  & $188\,774$ &  &  & $1\,285\,568$ \\
Log-likelihood &  &  & $-784232$ &  &  & $-5482604$ \\
 \\
\end{tabular}
\caption{Message length by desirability gap.\label{tab:supplement3}}
\end{center}
\end{table*}

\begin{table*}
\begin{center}
\begin{tabular}{lrrrrrrr}
 & & \multicolumn{3}{c}{Women} & \multicolumn{3}{c}{Men} \\
 & & Coef. & Std. error & $z$ & Coef. & Std. error & $z$ \\
 & & \multicolumn{3}{c}{\hrulefill} & \multicolumn{3}{c}{\hrulefill} \\
constant &  & $-2.579$ & $0.016$ & $-165.5$ & $-2.442$ & $0.006$  & $393.3$ \\
Chicago &  & $0.123$ & $0.017$ & $7.030$ & $0.053$ & $0.007$ & $7.890$ \\
NYC &  & $0.043$ & $0.016$ & $2.590$ & $0.014$ & $0.007$ & $2.190$ \\
Seattle &  & $0.113$ & $0.020$ & $5.750$ & $-0.014$ & $0.008$ & $-1.810$ \\
 \\
Desirability Gap &  & $0.134$ & $0.035$ & $3.820$ & $-0.102$ & $0.013$ & $-7.670$ \\
Desirability Gap $\times$ Chicago &  & $-0.022$ & $0.040$ & $-0.550$ & $-0.127$ & $0.015$ & $-8.780$ \\
Desirability Gap $\times$ NYC &  & $0.023$ & $0.037$ & $0.620$ & $-0.055$ & $0.014$ & $-3.940$ \\
Desirability Gap $\times$ Seattle &  & $0.004$ & $0.045$ & $0.100$ & $-0.011$ & $0.017$ & $-0.670$ \\
 \\
Desirability Gap$^2$ &  & $-0.001$ & $0.067$ & $-0.010$ & $-0.133$ & $0.027$ & $-5.010$ \\
Desirability Gap$^2$ $\times$ Chicago &  & $0.072$ & $0.076$ & $0.940$ & $0.197$ & $0.029$ & $6.850$ \\
Desirability Gap$^2$ $\times$ NYC &  & $-0.024$ & $0.071$ & $-0.330$ & $0.259$ & $0.028$ & $9.290$ \\
Desirability Gap$^2$ $\times$ Seattle &  & $0.038$ & $0.086$ & $0.440$ & $0.133$ & $0.033$ & $4.010$ \\
 \\
$N$  &  &  & $188\,774$ &  &  & $1\,285\,568$ \\
Log Likelihood &  &  & $-50732$ &  &  & $-367246$ \\
 \\
\end{tabular}
\caption{Proportion of positive words in message by desirability gap.\label{tab:supplement4}}
\end{center}
\end{table*}

\begin{table*}
\begin{center}
\begin{tabular}{lrrrrrrr}
 & & Women &  &  & Men \\
 & Coef. & Std. error & $z$ & Coef. & Std. error & $z$ \\
 & \multicolumn{3}{c}{\hrulefill} & \multicolumn{3}{c}{\hrulefill} \\
constant & $-0.130$ & $0.031$ & $-4.170$ & $-1.276$ & $0.014$ & $-91.220$ \\
Chicago & $-0.077$ & $0.034$ & $-2.270$ & $-0.053$ & $0.015$ & $-3.490$ \\
NYC & $-0.321$ & $0.033$ & $-9.770$ & $-0.218$ & $0.015$ & $-14.770$ \\
Seattle & $0.234$ & $0.040$ & $5.880$ & $0.046$ & $0.018$ & $2.570$ \\
 \\
Desirability gap & $-0.735$ & $0.051$ & $-14.410$ & $-0.500$ & $0.023$ & $-21.390$ \\
Desirability gap $\times$ Chicago & $0.104$ & $0.057$ & $1.830$ & $0.116$ & $0.026$ & $4.530$ \\
Desirability gap $\times$ NYC & $0.019$ & $0.054$ & $0.350$ & $0.025$ & $0.025$ & $1.020$ \\
Desirability gap $\times$ Seattle & $-0.030$ & $0.068$ & $-0.450$ & $0.210$ & $0.030$ & $6.940$ \\
 \\
Desirability gap$^2$  & $0.226$ & $0.101$ & $2.240$ & $0.285$ & $0.045$ & $6.260$ \\
Desirability gap$^2$ $\times$ Chicago & $0.213$ & $0.112$ & $1.900$ & $-0.283$ & $0.050$ & $-5.660$ \\
Desirability gap$^2$ $\times$ NYC & $0.280$ & $0.107$ & $2.620$ & $-0.293$ & $0.048$ & $-6.060$ \\
Desirability gap$^2$ $\times$ Seattle & $0.100$ & $0.134$ & $0.750$ & $-0.212$ & $0.058$ & $-3.640$ \\
 \\
Number of words & $-0.140$ & $0.111$ & $-1.260$ & $-0.059$ & $0.041$ & $-1.440$ \\
Number of words $\times$ Chicago & $0.279$ & $0.118$ & $2.370$ & $-0.040$ & $0.044$ & $-0.900$ \\
Number of words $\times$ NYC & $0.111$ & $0.117$ & $0.950$ & $0.054$ & $0.042$ & $1.270$ \\
Number of words $\times$ Seattle & $0.271$ & $0.125$ & $2.170$ & $0.243$ & $0.047$ & $5.210$ \\
 \\
Number of words$^2$ & $0.087$ & $0.058$ & $1.500$ & $0.035$ & $0.015$ & $2.310$ \\
Number of words$^2$ $\times$ Chicago & $-0.086$ & $0.059$ & $-1.450$ & $0.001$ & $0.016$ & $0.060$ \\
Number of words$^2$ $\times$ NYC & $-0.002$ & $0.061$ & $-0.040$ & $-0.036$ & $0.015$ & $-2.350$ \\
Number of words$^2$ $\times$ Seattle & $-0.078$ & $0.060$ & $-1.300$ & $-0.039$ & $0.016$ & $-2.460$ \\
 \\
$N$ &  & $188\,774$ &  &  & $1\,285\,568$ \\
Log-likelihood &  & $-126\,679$ &  &  & $-637\,918$ \\
 \\
\end{tabular}
\caption{Probability of reply by message length, conditional on desirability gap.\label{tab:supplement5}}
\end{center}
\end{table*}

\begin{table*}
\begin{center}
\begin{tabular}{lrrrrrrr}
 & & Women &  &  & Men \\
 & Coef. & Std. error & $z$ & Coef. & Std. error & $z$ \\
 & \multicolumn{3}{c}{\hrulefill} & \multicolumn{3}{c}{\hrulefill} \\
constant & $-0.167$ & $0.030$ & $-5.550$ \\
Chicago & $0.009$ & $0.033$ & $0.290$ & $-0.057$ & $0.016$ & $-3.610$ \\
NYC & $-0.288$ & $0.032$ & $-9.110$ & $-0.223$ & $0.015$ & $-14.660$ \\
Seattle & $0.374$ & $0.040$ & $9.320$ & $0.125$ & $0.019$ & $6.610$ \\
 \\
Desirability gap & $-0.742$ & $0.051$ & $-14.580$ & $-0.503$ & $0.023$ & $-21.540$ \\
Desirability gap $\times$ Chicago & $0.124$ & $0.057$ & $2.180$ & $0.111$ & $0.026$ & $4.320$ \\
Desirability gap $\times$ NYC & $0.029$ & $0.054$ & $0.540$ & $0.025$ & $0.025$ & $1.000$ \\
Desirability gap $\times$ Seattle & $-0.006$ & $0.068$ & $-0.090$ & $0.223$ & $0.030$ & $7.380$ \\
 \\
Desirability gap$^2$ & $0.227$ & $0.101$ & $2.250$ & $0.279$ & $0.045$ & $6.150$ \\
Desirability gap$^2$ $\times$ Chicago & $0.207$ & $0.112$ & $1.850$ & $-0.276$ & $0.050$ & $-5.520$ \\
Desirability gap$^2$ $\times$ NYC & $0.278$ & $0.107$ & $2.600$ & $-0.285$ & $0.048$ & $-5.890$ \\
Desirability gap$^2$ $\times$ Seattle & $0.096$ & $0.133$ & $0.720$ & $-0.213$ & $0.058$ & $-3.660$ \\
 \\
\% Positive words  & $0.003$ & $0.004$ & $0.850$ & $-0.006$ & $0.002$ & $-3.830$ \\
\% Positive words $\times$ Chicago & $-0.005$ & $0.004$ & $-1.320$ & $-0.001$ & $0.002$ & $-0.870$ \\
\% Positive words $\times$ NYC & $-0.003$ & $0.004$ & $-0.910$ & $0.002$ & $0.002$ & $1.220$ \\
\% Positive words $\times$ Seattle & $-0.013$ & $0.005$ & $-2.700$ & $-0.001$ & $0.002$ & $-0.230$ \\
 \\
\% Positive words$^2$  & $0.000$ & $0.000$ & $-0.610$ & $0.000$ & $0.000$ & $1.690$ \\
\% Positive words$^2$ $\times$ Chicago & $0.000$ & $0.000$ & $0.430$ & $0.000$ & $0.000$ & $1.540$ \\
\% Positive words$^2$ $\times$ NYC & $0.000$ & $0.000$ & $1.480$ & $0.000$ & $0.000$ & $-0.670$ \\
\% Positive words$^2$ $\times$ Seattle & $0.000$ & $0.000$ & $1.510$ & $0.000$ & $0.000$ & $-1.070$ \\
 \\
$N$ &  & $188\,774$ &  &  & $1\,285\,568$ \\
Log Likelihood &  & $-126\,705$ &  &  & $-637\,843$ \\
 \\
\end{tabular}
\caption{Probability of reply by percent of positive words, conditional on desirability gap.\label{tab:supplement6}}
\end{center}
\end{table*}


\begin{thebibliography}{20}
\bibitem{note1}
A tendency for messages to go to more desirable people is to some extent implicit in the PageRank measure, which often (though not always) rates people who receive a lot of messages as desirable.  However, the details of the distribution, including modal value, skewness, consistency across cities, and difference between women and men, are by no means inevitable and contain real information about partner choice and attraction.
\bibitem{note4}
Online dating site usage varies among demographic groups.  Being white, highly educated, high income, or having a school-aged child in the household are all positive predictors of internet use.  Nevertheless, as internet use and online dating becomes more commonplace, online populations increasingly resemble the broader US population (though there are some persistent differences).
\end{thebibliography}
\end{document}